\documentclass[12pt,a4paper]{article}
\usepackage{amssymb,amsfonts,amsthm,amsmath}

\usepackage{epsf}
\usepackage{graphicx}
\usepackage{placeins}
\usepackage{setspace}

\def\hang{\hangindent\parindent}
 \def\rf{\par\noindent\hang}

\newtheorem{theorem}{Theorem}

\newtheorem{lemma}{Lemma}
\newtheorem{corollary}{Corollary}

\begin{document}
\baselineskip=20pt

\begin{center} \Large{{\bf FURTHER PROPERTIES OF FREQUENTIST CONFIDENCE
INTERVALS IN REGRESSION THAT UTILIZE UNCERTAIN PRIOR INFORMATION}}
\end{center}

\bigskip

Running title: CONFIDENCE INTERVALS AND PRIOR INFORMATION

\bigskip

\begin{center}
\large{
{\bf PAUL KABAILA$^{1*}$ {\normalsize \rm AND} KHAGESWOR GIRI$^2$}}
\end{center}

\medskip

\begin{center}
{\large{\sl La Trobe University and Victorian Department of Primary Industries}}
\end{center}

\vspace{11cm}

\noindent * Author to whom correspondence should be addressed.

\bigskip

\noindent 1. Department of Mathematics and Statistics,
La Trobe University, Victoria 3086, Australia. e-mail: P.Kabaila@latrobe.edu.au. Facsimile: +61 3 9479 2466.
Telephone: +61 3 9479 2594.

\bigskip

\noindent 2. Department of Primary Industries, 600 Sneydes Road Werribee, Victoria 3030, Australia. e-mail: khageswor.giri@dpi.vic.gov.au

\newpage

%\medskip

\begin{center}
{\bf Summary}
\end{center}

%\medskip

\noindent Consider a linear regression model with $n$-dimensional response vector,
regression parameter $\boldsymbol{\beta} = (\beta_1, \ldots,  \beta_p)$ and
independent and identically $N(0, \sigma^2)$ distributed errors.
Suppose that the parameter of interest is $\theta = \boldsymbol{a}^T \boldsymbol{\beta}$
where $\boldsymbol{a}$ is a specified vector.
Define the parameter $\tau = \boldsymbol{c}^T \boldsymbol{\beta} - t$ where  $\boldsymbol{c}$ and $t$
are specified.
Also suppose that we have uncertain prior information that $\tau = 0$.
Part of our evaluation of a frequentist confidence interval for $\theta$ is the ratio
(expected length of this confidence interval)/(expected length of standard $1-\alpha$ confidence interval),
which we call the scaled expected length of this interval.
We say that a $1-\alpha$ confidence interval for $\theta$ utilizes this
uncertain prior information if (a) the scaled expected length of this interval is significantly
less than 1 when $\tau = 0$, (b) the maximum value
of the scaled expected length is not too much larger than 1 and (c) this
confidence interval reverts to the standard $1-\alpha$ confidence interval
when the data happen to strongly contradict the prior information. Kabaila \& Giri, 2009,
{\sl JSPI} present a new method for finding such a
confidence interval. Let $\hat{\boldsymbol{\beta}}$ denote the least squares estimator of $\boldsymbol{\beta}$.
Also let $\hat \Theta=\boldsymbol{a}^T \hat{\boldsymbol{\beta}}$ and
$\hat \tau=\boldsymbol{c}^T \hat{\boldsymbol{\beta}} - t$.
Using computations and new theoretical results, we show that the performance of this
confidence interval improves as $|\text{Corr}(\hat \Theta, \hat \tau)|$ increases and $n-p$ decreases.

\bigskip
%\medskip


\vspace{0.7cm}

\newpage

\begin{center}
{\bf 1. Introduction}
\end{center}

\medskip

Hodges \& Lehmann (1952), Bickel (1984) and Kempthorne (1983, 1987, 1988) present frameworks
for the utilization of uncertain prior information (about the parameters of the model)
in frequentist inference, mostly for point estimation. Such information can arise from previous experience with similar data sets and/or
expert opinion and scientific background.
We say that the confidence set ${\cal C}$ is a $1-\alpha$ confidence
set for the parameter of interest $\theta$ if its infimum coverage probability is the specified value $1-\alpha$. We assess such a confidence
set by its scaled expected volume, defined to be the ratio (expected volume of ${\cal C}$)/(expected
volume of the standard $1-\alpha$ confidence set).
The first requirement of a
$1-\alpha$ confidence set that utilizes the uncertain
prior information is that its scaled expected volume is significantly less than 1 when the
prior information is correct (Kabaila, 2009).

Confidence sets that satisfy this first requirement can be classified
into the following two groups. The first group consists of
$1-\alpha$ confidence sets with scaled expected volume
that is less than or equal to 1 for all parameter values, so that these dominate the
standard $1-\alpha$ confidence set. Examples of such confidence sets are the Stein-type confidence interval
for the normal variance (see e.g. Maata \& Casella, 1990 and Goutis \& Casella, 1991) and Stein-type confidence sets for the multivariate
normal mean (see e.g. Stein, 1962, Berger, 1980, Casella \& Hwang, 1983,
Tseng \& Brown, 1997, Efron, 2006 and Saleh, 2006). The second group consists of $1-\alpha$
confidence sets that satisfy this first requirement, when dominance of the usual $1-\alpha$ confidence set is not possible (the
scaled expected volume must exceed 1 for some parameter values). This second
group includes confidence intervals described by Pratt (1961), Brown et al (1995)
and Puza \& O'Neill (2006ab). This second group also includes
$1-\alpha$ confidence sets that satisfy the additional requirements that (a) the maximum (over the parameter space)
of the scaled expected volume is not too much larger than 1 and (b) the confidence set reverts to
the usual $1-\alpha$ confidence set when the data happen to strongly contradict the prior information.
Confidence intervals that utilize uncertain prior information and satisfy these additional requirements have been proposed by Farchione \& Kabaila (2008)
and Kabaila \& Giri (2009). The purpose
of the present paper is to analyse further interesting properties of the Kabaila \& Giri (2009) confidence interval.

 Consider the linear regression model $\boldsymbol{Y} = \boldsymbol{X} \boldsymbol{\beta} + \boldsymbol{\varepsilon}$,
 where $Y$ is a random $n$-vector of responses, $\boldsymbol{X}$ is a known $n \times p$ matrix with linearly
independent columns, $\boldsymbol{\beta} = (\beta_1,\ldots, \beta_p)$ is an unknown parameter vector and
$\varepsilon \sim N(0, \sigma^2 I_n)$ where $\sigma^2$ is an unknown positive parameter.
Suppose that the parameter of interest is $\theta = \boldsymbol{a}^T \boldsymbol{\beta}$
where $\boldsymbol{a}$ is specified
$p$-vector ($\boldsymbol{a} \ne \boldsymbol{0}$).
The inference of interest is a $1-\alpha$ confidence interval for $\theta$.
Define the parameter $\tau = \boldsymbol{c}^T \boldsymbol{\beta} - t$ where the vector $\boldsymbol{c}$ and the number $t$
are specified and $\boldsymbol{a}$ and $\boldsymbol{c}$ are linearly independent.
Also suppose that we have uncertain prior information that $\tau = 0$.

Part of our evaluation of a frequentist confidence interval for $\theta$ is the ratio
\begin{equation*}
\frac{(\text{expected length of this confidence interval})}{
(\text{expected length of standard } 1-\alpha \text{ confidence interval})},
\end{equation*}
where the standard $1-\alpha$ confidence interval is obtained by fitting the
full model to the data.
We call this ratio the scaled expected length of this confidence interval.
We say that a $1-\alpha$ confidence interval for $\theta$ utilizes this
uncertain prior information if the following three conditions hold.
The first condition is that the scaled expected length of this interval is significantly
less than 1 when $\tau = 0$. The strong admissibility of the standard $1-\alpha$ confidence interval, as proved by
Kabaila, Giri \& Leeb (2010), implies that the maximum (over the parameter space)
of the scaled expected length of this interval must be greater than 1.
The second condition is that this maximum
is not too much larger than 1. The third condition is that this
confidence interval reverts to the standard $1-\alpha$ confidence interval
when the data happen to strongly contradict the uncertain prior information
that $\tau = 0$.

Kabaila and Giri (2009) present a new method for finding such a
confidence interval. For convenience, we refer to the confidence interval
found by this method as the KG confidence interval.
This method is described briefly in the next section.
Let $\hat{\boldsymbol{\beta}}$ denote the least squares estimator of $\boldsymbol{\beta}$.
Also let $\hat \Theta$ denote $\boldsymbol{a}^T \hat{\boldsymbol{\beta}}$ and $\hat \tau$ denote
$\boldsymbol{c}^T \hat{\boldsymbol{\beta}} - t$.
We elucidate the dependence of the properties of this confidence
interval on Corr$(\hat \Theta, \hat \tau)$ and $n-p$. Note that Corr$(\hat \Theta, \hat \tau)$
is determined by $\boldsymbol{a}$, $\boldsymbol{c}$ and $\boldsymbol{X}$,
so that it does not depend on the unknown parameters
$\boldsymbol{\beta}$ and $\sigma^2$.

In Section 3, we consider the dependence of these properties on $n-p$, when
Corr$(\hat \Theta, \hat \tau)=0$. We prove that the KG confidence interval
is centred at $\hat \Theta$ and is equi-tailed.
Using computations and a new theoretical result, we show that
that the KG confidence interval (a) utilizes the uncertain prior information
for small $n-p$ and (b) loses the ability to utilize this uncertain prior information as $n-p$
increases.
Let $\hat{\sigma}^2$ denote the usual unbiased estimator of $\sigma^2$, obtained by fitting
the full model.
Our explanation for this finding is that when Corr$(\hat \Theta, \hat \tau)=0$, the ability of the
KG confidence interval to utilize the uncertain prior information
comes from the ability to estimate $\sigma^2$ with greater accuracy than by using $\hat \sigma^2$,
particularly when $n-p$ is small.
%This finding is consistent with
%the existence of improved confidence intervals for the normal %variance,
%see e.g. Goutis \& Casella (1991).

In Section 4, we consider the dependence of the properties of the KG interval on $n-p$, when
Corr$(\hat \Theta, \hat \tau) \ne 0$. We show, through computational results,
that the KG confidence interval utilizes the uncertain prior information
irrespective of how large $n-p$ is, with increasing ability to do so
when $|$Corr$(\hat \Theta, \hat \tau)|$ is large. Our interpretation of this
finding is that Corr$(\hat \Theta, \hat \tau) \ne 0$ provides another source
of the ability to utilize the uncertain prior information.

Our overall conclusion is that there are two sources
for the ability of a $1-\alpha$ confidence interval for $\theta$ to utilize the uncertain prior information.
The first of these sources is a non-zero Corr$(\hat \Theta, \hat \tau)$. The second of these sources is the
ability, for small and medium $n-p$, to estimate $\sigma^2$ with more accuracy.
The performance of the KG
confidence interval improves as $|\text{Corr}(\hat \Theta, \hat \tau)|$ increases and $n-p$ decreases.

The scaled expected length of the KG interval is a function
of the parameter $\gamma = \tau/\sqrt{\text{Var}(\hat{\tau})}$.
Figure 2 is a plot of the squared scaled expected length (which is an even function of $\gamma$)
as a function of $\gamma$ for
this interval, with tuning parameter $\xi = 0.15$, for the case that Corr$(\hat \Theta, \hat \tau) = 0.8165$,
$n-p=1$ and $1-\alpha=0.95$. When the prior information is correct (i.e. when $\gamma=0$), we gain a great deal since the
squared scaled expected length is 0.6960. The maximum value of the squared scaled expected length is only 1.0626.
This confidence interval reverts to the standard $1-\alpha$ confidence interval when the data strongly
contradict the uncertain prior information that $\tau=0$. This is reflected by the fact that the squared scaled expected length
converges to 1 as $\gamma \rightarrow \infty$.

%\newpage
\baselineskip=19pt

\medskip

\begin{center}
{\bf 2. Description of the confidence interval of Kabaila \& Giri (2009)}
\end{center}

%\medskip

Let $v_{11} = \text{Var}(\hat{\Theta})/\sigma^2$, $v_{22} = \text{Var}(\hat{\tau})/\sigma^2$
and $v_{12} = \text{Cov}(\hat \Theta, \hat \tau)/\sigma^2$
The standard $1-\alpha$ confidence interval for $\theta$ is
$I = \big [ \hat \Theta - t(n-p) \sqrt{v_{11}} \hat
\sigma, \quad \hat \Theta + t(n-p) \sqrt{v_{11}} \hat
\sigma \big ]$,
where the quantile $t(m)$ is defined by $P(-t(m) \le T \le t(m)) = 1-\alpha$ for $T \sim t_m$ and
$\hat \sigma^2 = (\boldsymbol{Y} - \boldsymbol{X} \hat{\boldsymbol{\beta}})^T (\boldsymbol{Y} - \boldsymbol{X} \hat{\boldsymbol{\beta}})/(n-p)$.

Henceforth, suppose that $b: \mathbb{R} \rightarrow \mathbb{R}$ is an odd function and
$s: [0, \infty) \rightarrow (0, \infty)$ are measurable functions.
We use the notation $[\tilde{a} \pm \tilde{b}]$ for the interval $[\tilde{a}-\tilde{b}, \tilde{a}+\tilde{b}]$
($\tilde{b}>0$).
For each $b$ and $s$,
define the following confidence interval for $\theta$
\begin{align*}
%\label{J(b,s)}
J(b,  s) = \bigg [ \hat \Theta -
\sqrt{v_{11}} \hat \sigma \, b\bigg(\frac{\hat{\tau}}{\hat \sigma \sqrt{v_{22}}}\bigg) \, \pm \,
\sqrt{v_{11}} \hat \sigma \, s\bigg(\frac{|\hat{\tau}|}{\hat \sigma \sqrt{v_{22}}}\bigg)
\bigg ].
\end{align*}
Let $\gamma = \tau/\sqrt{\text{Var}(\hat{\tau})} = \tau/(\sigma \sqrt{v_{22}})$ and
$\rho = \text{Corr}(\hat \Theta, \hat \tau)= v_{12}/\sqrt{v_{11} v_{22}}$. Note that $\rho=\big(\boldsymbol{a}^T (\boldsymbol{X}^T \boldsymbol{X})^{-1}\boldsymbol{c} \big)/ \sqrt{\boldsymbol{a}^T (\boldsymbol{X}^T \boldsymbol{X})^{-1}\boldsymbol{a} \, \boldsymbol{c}^T (\boldsymbol{X}^T \boldsymbol{X})^{-1}\boldsymbol{c}} \,$ and so does not depend on the unknown parameters
$\boldsymbol{\beta}$ and $\sigma^2$.
For given $(b,s, \rho)$,
the coverage probability $P\big( \theta \in J(b, s) \big)$ is an even function of $\gamma$,
which we denote by $c(\gamma;b,s, \rho)$.
The scaled expected length of $J(b,s)$ is (expected length of $J(b, s)$)/(expected length of $I$)
and is an even function of $\gamma$ for given $s$, which we denote
by $e(\gamma;s)$.

Define the class ${\cal B}$ to consist of the odd functions $b: \mathbb{R} \rightarrow \mathbb{R}$ that
satisfy $b(x)=0$ for all $|x| \ge d$, where $d$ is a (sufficiently large) specified positive number.
Also define the class ${\cal S}$ to consist of the functions $s: [0, \infty)
\rightarrow (0, \infty)$, where $s(x)=t(n-p)$ for all $x \ge d$.
Stated briefly, we find the $1-\alpha$ confidence interval for $\theta$ that utilizes the uncertain
prior information that $\tau = 0$ as follows. Find smooth functions $b \in {\cal B}$ and $s \in {\cal S}$
such that (a) the minimum of $c(\gamma;b,s, \rho)$ over $\gamma$ is
$1-\alpha$ and (b)
\begin{equation}
\label{criterion}
\xi \int_{-\infty}^{\infty} (e(\gamma;s) - 1) \, d \gamma + (e(0;s)-1)
\end{equation}
is minimized, where $\xi$ is a specified nonnegative tuning parameter.
The larger the value of $\xi$, the smaller the relative weight given to
minimizing $e(\gamma;s)$ for $\gamma=0$, as opposed to
minimizing $e(\gamma;s)$ for other values of $\gamma$.
Since we require that
$b \in {\cal B}$ and $s \in {\cal S}$, this confidence interval reverts to the standard $1-\alpha$ confidence interval
$I$ when the data happen to strongly contradict the uncertain prior information that $\tau = 0$.
The tuning parameter $\xi$ and
the functions $b$ and $s$ are chosen by the statistician {\sl prior} to looking at the observed
response vector $\boldsymbol{y}$. Further details
of the method used to make this choice are provided in Appendix A.

%\newpage

\medskip

\noindent \textbf{Example 1 ($\boldsymbol{2^3}$ factorial experiment without replication)} \newline
Consider a $2^3$ factorial experiment without replication. Let $Y$ denote the response and let
$x_1$, $x_2$ and $x_3$ denote the coded levels for each of the 3 factors, where the coded level
takes either the value $-1$ or 1. We will assume the model
\begin{equation*}
Y=\beta_{0}+\beta_{1}x_{1} +\beta_{2}x_{2}+\beta_{3}x_{3}+\beta_{12}x_{1}x_{2}
+\beta_{13}x_{1}x_{3}+\beta_{23}x_{2}x_{3}+\beta_{123}x_{1}x_{2}x_{3}+ \varepsilon
\end{equation*}
where  $\beta_{0}$,
$\beta_{1}$, $\beta_{2}$, $\beta_{3}$, $\beta_{12}$, $\beta_{13}$,
$\beta_{23}$, $\beta_{123}$ are unknown parameters and
$\varepsilon \sim N(0, \sigma^2 )$, where $\sigma^2$ is an unknown
positive parameter.

For factorial experiments it is
commonly believed that higher order interactions are negligible
(see e.g. Mead (1988, p.368) and Hinkelman \& Kempthorne (1994, p.350)). Indeed, this type of
belief is the basis for the design of fractional factorial experiments.
Suppose that $\beta_{123} = 0$ and that we have uncertain prior information that
$\beta_{12}$, $\beta_{13}$ and $\beta_{23}$ are all zero. Thus $n-p = 1$.
We consider the particular case that the parameter of interest
interest $\theta$ is the contrast $\big(E(Y) \text{ for } (x_1,x_2,x_3)=(1,-1,-1) \big) -
\big(E(Y) \text{ for } (x_1,x_2,x_3)=(1,-1,1) \big)$.
%
%\begin{equation*}
%\big(E(Y) \text{ for } (x_1,x_2,x_3)=(1,-1,-1) \big) -
%\big(E(Y) \text{ for } (x_1,x_2,x_3)=(1,-1,1) \big).
%\end{equation*}
%
In other words, $\theta =
2\beta_{123}-2\beta_{13}+2\beta_{23}-2\beta_{3}$. Since we assume that $\beta_{123} = 0$,
$\theta = -2\beta_{13}+2\beta_{23}-2\beta_{3}$.

Let $\tau = \beta_{23} - \beta_{13}$. The uncertain
prior information that $\beta_{12}$, $\beta_{13}$ and $\beta_{23}$ are all zero implies the
uncertain prior information that $\tau = 0$.
Note that Corr$(\hat{\Theta}, \hat{\tau}) = \sqrt{2/3} =
0.816496$.
Figure 1 is a plot of the functions $b$ and $s$ for the KG $1-\alpha$ confidence interval
for $\theta$ when Corr$(\hat{\Theta}, \hat{\tau}) = 0.816496$, $n-p=1$, $1-\alpha = 0.95$, $\xi = 0.15$, $d = 40$,
the knots of the cubic spline $b$ (in the interval $[0,d]$) at $0, 15, 18, 21, 24, 27, 30, 40$ and
the knots of the cubic spline $s$ (in the interval $[0,d]$) at  $0, 3, 6, 9, 12, 15, 30, 40$.
To an excellent approximation, the coverage probability of this confidence interval is 0.95
for all $\gamma$. The minimum coverage probability of this confidence interval is 0.94992.
Figure 2 is a plot of the squared scaled expected length of this confidence interval
as a function of $\gamma$. When the prior information is correct (i.e. when $\gamma=0$),
we gain a great deal since the squared scaled expected length is 0.6960.
For $\gamma$ larger than 15, the squared scaled expected length is a decreasing function
and approaches 1 as $\gamma \rightarrow \infty$.

\baselineskip=20pt

% The following command was used to prevent the figures moving around too much.
\FloatBarrier

\begin{figure}[h]
\label{Figure1}
%\vskip-3cm
    %\includegraphics[keepaspectratio=true, width=15cm]{fig1_vilnius_13jun06}
    \hskip-1cm
    %\centering
    \includegraphics[scale=0.4]{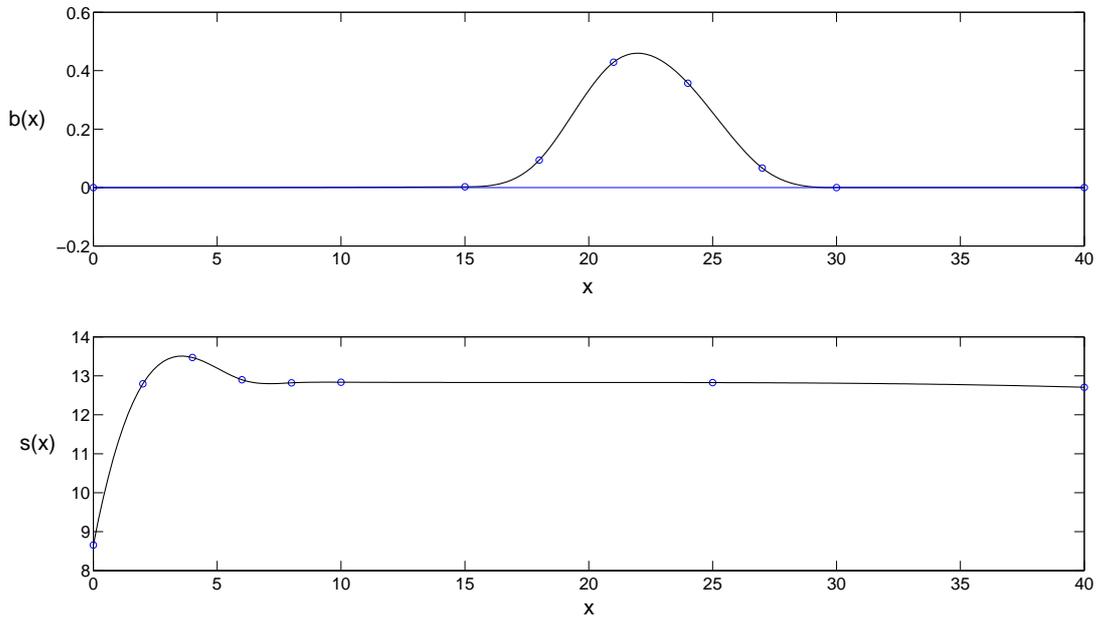}
    \caption{Plots of the functions $b$ and $s$ for the KG $1-\alpha$ confidence interval
for $\theta$ when $\text{Corr}(\hat{\Theta}, \hat{\tau}) =
0.816496$, $n-p=1$, $1-\alpha = 0.95$, $\xi = 0.15$, $d = 40$ and
the knots of the cubic splines $b$  and $s$ (in the interval $[0,d]$) are at $0, 15, 18, 21, 24, 27, 30, 40$ and
 at  $0, 2, 4, 6, 8, 10, 25, 40$, respectively.}
\end{figure}

\FloatBarrier

% The following command was used to prevent the figures moving around too much.
\FloatBarrier

\begin{figure}[h]
\label{Figure2}
%\vskip-3cm
    %\includegraphics[keepaspectratio=true, width=15cm]{fig1_vilnius_13jun06}
    \hskip-0.5cm
    %\centering
    \includegraphics[scale=0.38]{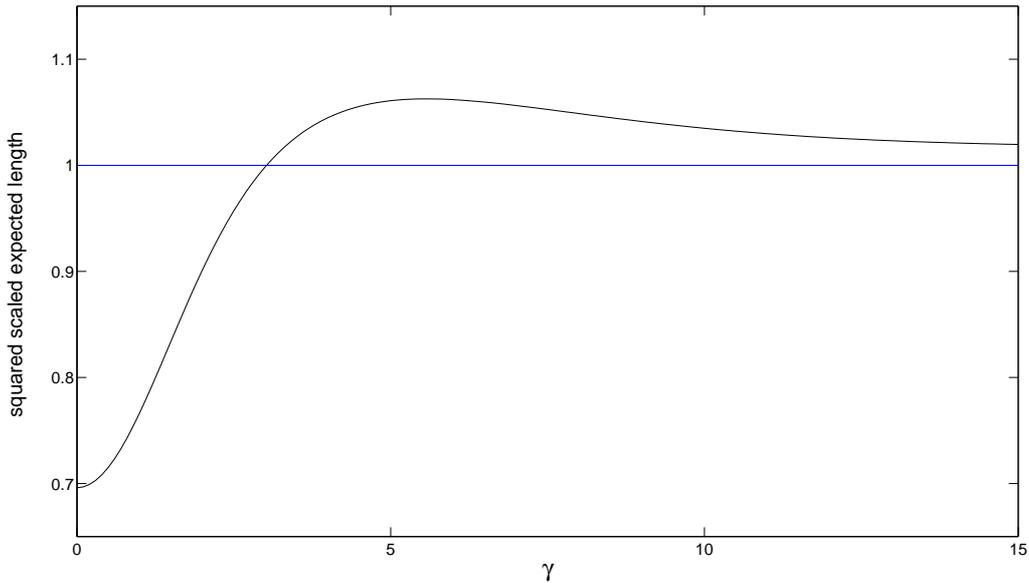}
    \caption{Plot of the squared scaled expected length $e^2(\gamma;s)$ (as a function of
   $\gamma = \tau/(\sigma \sqrt{v_{22}})$) for the KG $1-\alpha$ confidence interval
for $\theta$ when $\text{Corr}(\hat{\Theta}, \hat{\tau}) =
0.816496$, $n-p=1$, $1-\alpha = 0.95$, $\xi = 0.15$, $d = 40$ and
the knots of the cubic splines $b$  and $s$ (in the interval $[0,d]$) are at $0, 15, 18, 21, 24, 27, 30, 40$ and
 at  $0, 2, 4, 6, 8, 10, 25, 40$, respectively.}
\end{figure}

\FloatBarrier

\medskip

\begin{center}
{\bf 3. Performance of the KG interval
%as a function of $\boldsymbol{n-p}$
for Corr$\boldsymbol{(\hat{\Theta}, \hat{\tau})=0}$}
\end{center}

%\medskip

In this section we consider the case that Corr$(\hat \Theta, \hat \tau)=0$. For notational convenience,
we use $b \equiv 0$ to denote the function $b: \mathbb{R} \rightarrow \mathbb{R}$ satisfying
$b(x)=0$ for all $x \in \mathbb{R}$.
Corollary 1 (stated later in this section) shows that
choosing $b \equiv 0$ does not lead to any loss in the performance of the KG confidence
interval for $\theta$. We therefore make the restriction that $b \equiv 0$.
This implies that the KG confidence interval has the form
\begin{equation}
\label{ci_b_0}
\bigg [ \hat \Theta
\, \pm \,
\sqrt{v_{11}} \hat \sigma \, s\bigg(\frac{|\hat{\tau}|}{\hat \sigma \sqrt{v_{22}}}\bigg)
\bigg ],
\end{equation}
so that it is centred at $\hat{\Theta}$.
Theorem 2 shows
that the resulting KG confidence interval is equi-tailed.
As illustrated by Figure 3, computations show that the performance of
this confidence interval is good when $n-p$ is small, but degrades as $n-p$ increases and disappears as $n-p \rightarrow \infty$.

\FloatBarrier

\begin{figure}[h]
\label{Figure3}
%\vskip-3cm
    %\includegraphics[keepaspectratio=true, width=15cm]{fig1_vilnius_13jun06}
    \hskip-2cm
    %\centering
    \includegraphics[scale=0.45]{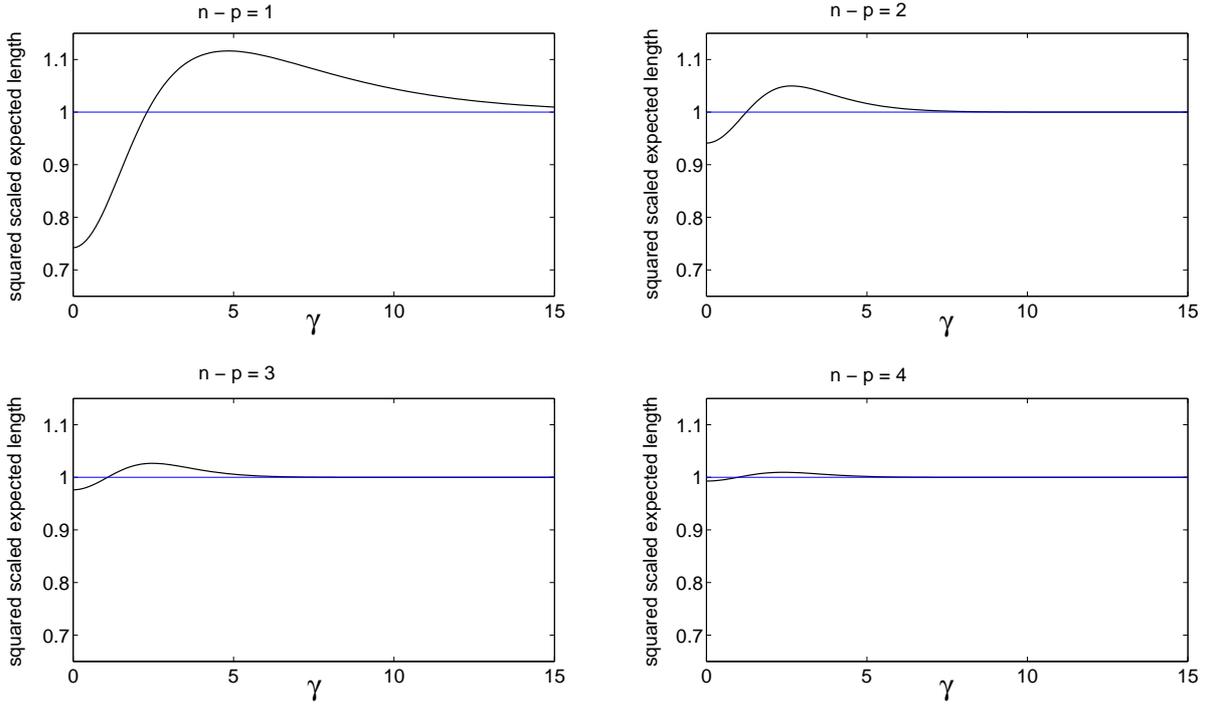}
    %\vskip-2.7cm
    \caption{Plots of the squared scaled expected length $e^2(\gamma;s)$ (as a function of
   $\gamma = \tau/(\sigma \sqrt{v_{22}})$) for the KG $1-\alpha$ confidence interval
for $\theta$ when $\text{Corr}(\hat{\Theta}, \hat{\tau}) =
0$, $1-\alpha = 0.95$, $\xi = 0.15$, $b \equiv 0$, $d = 12$ and
the knots of the cubic spline $s$ (in the interval $[0,d]$) are at $0, 1.5, 3, 4.5, 6, 7.5, 9, 10.5, 12$.
The values of $n-p$ are 1, 2, 3 and 4.}
\end{figure}
\FloatBarrier

\noindent
Theorem 3 proves the truth of this computational finding. The explanation for this finding is that
when Corr$(\hat \Theta, \hat \tau)=0$, the ability of the
KG confidence interval to utilize the uncertain prior information
comes from the ability to estimate $\sigma^2$ with greater accuracy than by using $\hat \sigma^2$.
This ability is significant when $n-p$ is small, but decreases as $n-p$ increases and disappears as $n-p \rightarrow \infty$.

The following theorem shows that for fixed function $s$, the coverage probability of the confidence interval
$J(b,s)$ is maximized by setting $b \equiv 0$.
\begin{theorem}
Suppose that Corr$(\hat \Theta, \hat \tau)=0$ and that the function $s \in {\cal S}$ is given.
For each $\gamma \in \mathbb{R}$, the coverage probability $c(\gamma;b,s,\rho)$ is maximized
with respect to the function $b \in {\cal B}$, by setting $b \equiv 0$.
\end{theorem}
\noindent This theorem is proved in Appendix C. The following result, which is a corollary of Theorem 1,
shows that choosing $b \equiv 0$
does not lead to any loss in the performance of the KG confidence
interval.

\newpage

\begin{corollary}
Suppose that Corr$(\hat \Theta, \hat \tau)=0$. Suppose that ${\cal B}^*$ is a subset of ${\cal B}$ that includes the function
$b \equiv 0$.
Also suppose that ${\cal S}^*$ is a subset of ${\cal S}$.
The infimum over $(b,s) \in {\cal B}^* \times {\cal S}^*$
of \eqref{criterion}, subject to the coverage constraint
\begin{equation}
\label{cov_constr}
c(\gamma;b,s, \rho)  \ge 1-\alpha \qquad \text{for all } \gamma \in \mathbb{R},
\end{equation}
is equal to the infimum over $s \in {\cal S}^*$ of \eqref{criterion}, subject to this constraint,
when $b \equiv 0$.
\end{corollary}
\noindent This corollary is proved in Appendix D.

The following theorem implies that if $b \equiv 0$ then the KG confidence interval is equi-tailed.
\begin{theorem}
Suppose that Corr$(\hat \Theta, \hat \tau)=0$ and that $b \equiv 0$.
Then the confidence interval $J(b,s)$ for $\theta$ is equi-tailed.
\end{theorem}
\noindent This theorem is proved in Appendix E. The following theorem shows that the performance of
this confidence interval degrades as $n-p$ increases and disappears as $n-p \rightarrow \infty$.
\begin{theorem}
Suppose that Corr$(\hat \Theta, \hat \tau)=0$ and that $b \equiv 0$.
Define
\begin{equation*}
\tilde{\cal S} = \big \{ s \in {\cal S}: c(\gamma;b,s, \rho) \ge 1-\alpha \text{ for all } \gamma \big \}.
\end{equation*}
Then
\begin{equation*}
\inf_{s \in \tilde{\cal S}} e(\gamma=0; s) \ge 1 - \eta_{n-p}
\end{equation*}
where $\{\eta_m\}$ is a sequence of positive numbers converging to 0 as
$m \rightarrow \infty$.
\end{theorem}
\noindent This theorem is proved in Appendix F. Although lengthy, this proof is quite
straightforward and elementary.

\medskip

\begin{center}
{\bf 4. Performance of the KG interval
for Corr$\boldsymbol{(\hat{\Theta}, \hat{\tau}) \ne 0}$}
\end{center}
%\medskip

In this section we consider the case that $\rho=\text{Corr}(\hat \Theta, \hat \tau) \ne 0$. For $n-p$ large,
$\hat \sigma^2$ estimates $\sigma^2$ with great accuracy and so the ability of the KG confidence interval
to utilize the uncertain prior information does not come from the estimation of $\sigma^2$ with more accuracy. This ability
comes instead from the correlation between $\hat \Theta$ and $\hat \tau$.
The computational results shown in Figure 4 for $n-p = 200$ illustrate this point well. For ease of comparison, Figures 2, 3 and 4
have the same limits on their horizontal and vertical axes.

\FloatBarrier

\begin{figure}[h]
%\label{Figure4}
%\vskip-3cm
    %\includegraphics[keepaspectratio=true, width=15cm]{fig1_vilnius_13jun06}
    \hskip-2cm
    %\centering
    \includegraphics[scale=0.45]{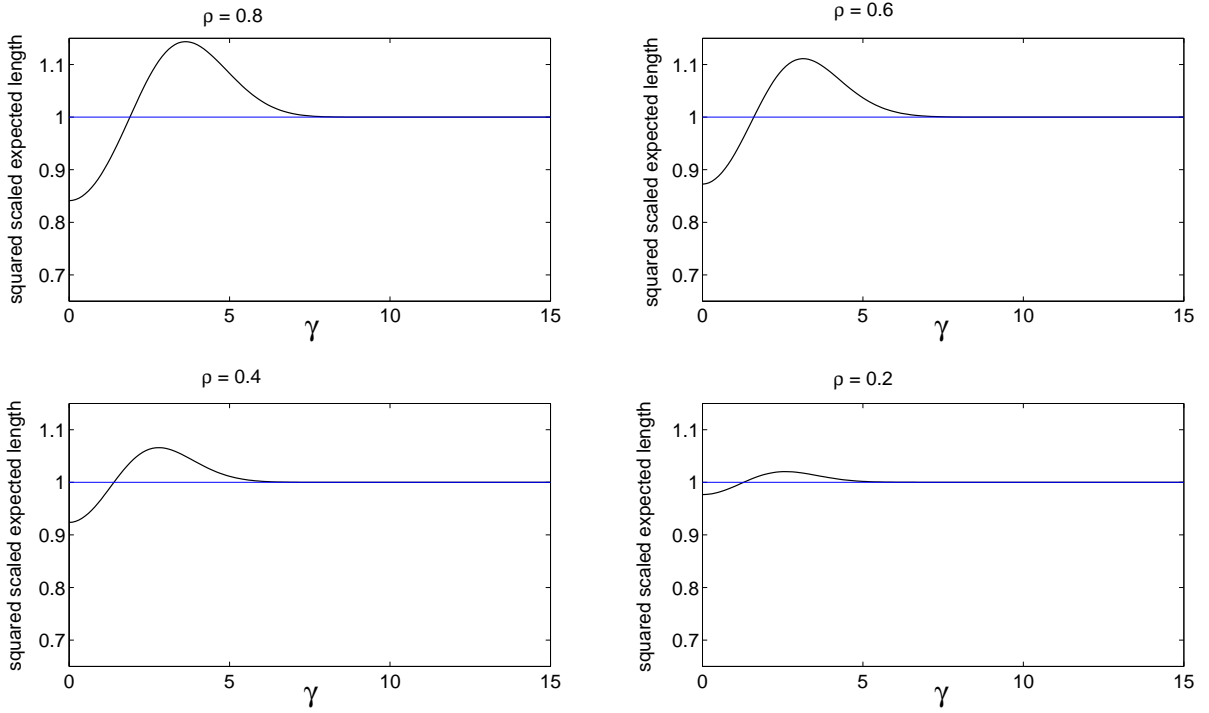}
    %\vskip-2.7cm
    \caption{Plots of the squared scaled expected length $e^2(\gamma;s)$ (as a function of
   $\gamma = \tau/(\sigma \sqrt{v_{22}})$) for the KG $1-\alpha$ confidence interval
for $\theta$ when $n-p=200$, $1-\alpha = 0.95$, $\xi = 0.15$, $d = 6$ and
the knots of the cubic splines $b$ and $s$ (in the interval $[0,d]$) are at $0, 1, 2, 3, 4, 5, 6$.
The values of $\rho = \text{Corr}(\hat{\Theta}, \hat{\tau})$ are 0.8, 0.6, 0.4 and 0.2.}
\end{figure}
\FloatBarrier

%\noindent A comparison of Figure 2 with the top left-hand panels of %Figures 3 and 4 suggests the following.
%The ability of the confidence interval to utilize prior information %that is displayed in Figure 2 has two
%important sources: (a) the ability to estimate $\sigma^2$ with greater %accuracy than by using $\hat{\sigma}^2$
%and (b) a large value of $|\text{Corr}(\hat \Theta, \hat \tau)|$.

\medskip

\begin{center}
\noindent {\bf 5. Remarks}
\end{center}

%\medskip

\noindent {\sl Remark 5.1} \ \ It might be hoped that a confidence interval constructed in the following way will be able
to utilize this uncertain prior information. Carry out a preliminary test of the null hypothesis
that $\tau = 0$ against the alternative hypothesis that $\tau \ne 0$. If this null hypothesis is rejected
then we use the standard $1-\alpha$ confidence interval for $\theta$. If, on the other hand, this
null hypothesis is accepted then we use the standard $1-\alpha$ confidence interval for $\theta$,
assuming that $\tau=0$. We call this the naive $1-\alpha$ confidence interval for $\theta$.
A computationally-convenient formula for the coverage probability of this confidence interval
is given in Theorem 3 of Kabaila \& Giri (2009b). The minimum coverage probability of this confidence
interval can be far below $1-\alpha$. Kabaila (1998) increases the half-width of this confidence interval,
when this null hypothesis is accepted, by the smallest possible value such that the adjusted interval
has minimum coverage $1-\alpha$. He shows that such confidence intervals can utilize the uncertain
prior information that $\tau=0$ when $n-p$ is small. However, this adjusted confidence interval
has the disadvantages that (a) it is obtained by an ad hoc adjustment, (b) there may be far better
adjustments and (c) the endpoints of this interval are discontinuous functions of the data.
Kabaila \& Giri (2009a) motivate the confidence interval
analysed in the present paper by greatly ``loosening up''
up the form of the naive $1-\alpha$ confidence interval
for $\theta$.

\medskip

\noindent {\sl Remark 5.2} \ \ If we knew (with certainty) that $\tau=0$ then the centre of the confidence interval for $\theta$
would be
\begin{equation}
\label{tau_known_0}
\hat \Theta -
\sqrt{v_{11}} \hat \sigma \, b\bigg(\frac{\hat{\tau}}{\hat \sigma \sqrt{v_{22}}}\bigg),
\end{equation}
where $b(x) = \rho x$. This fact provides a hint that the following results may be true:

\begin{enumerate}

\item[(R1)]

If $\rho=0$ then there is no loss in the performance of the KG interval if we make
the additional constraint that $b \equiv 0$.

\item[(R2)]

If $\rho > 0$ then there is no loss in the performance of the KG interval if we make
the additional constraint that $b \ge 0$ for all $x > 0$.

\item[(R3)]

If $\rho < 0$ then there is no loss in the performance of the KG interval if we make
the additional constraint that $b \le 0$ for all $x > 0$.

\end{enumerate}

\noindent As stated in Section 3 and proved in Appendix C, the result (R1) is true. Very
extensive numerical computations carried out by the authors suggest that the results (R2) and
(R3) are also true. For example, the top panel of Figure 1 of the present paper and the top panel of Figure 2 of
Kabaila \& Giri (2009a) are consistent with the results (R2) and
(R3), respectively. This strongly suggests that, for all possible data values, the centre of the KG interval
cannot be obtained by a shift from $\hat{\Theta}$ in the opposite direction to
\eqref{tau_known_0}.

\medskip

\noindent {\sl Remark 5.3} \ \
Suppose that we wish to construct an {\sl equi-tailed} $1-\alpha$ confidence interval for $\theta$ that utilizes
the available uncertain prior information.
As the following two examples show, consideration of the case that Corr$(\hat \Theta, \hat \tau)=0$ provides us with
a method of constructing such a confidence interval in the context of certain types of prior information.

\medskip

\noindent \textbf{Example 2 ($\boldsymbol{2^3}$ factorial experiment without replication, equi-tailed confidence interval
for $\boldsymbol{\theta}$)} \newline
Consider the same model, uncertain prior information and parameter of interest $\theta$ as delineated in the first two
paragraphs of the description of Example 1. Suppose that we wish to find an
{\sl equi-tailed} $1-\alpha$ confidence interval for $\theta$ that utilizes this prior information. We find such a confidence
interval by letting $\tau = \beta_{12}$. This uncertain prior information implies the uncertain prior information
that $\tau=0$. Note that Corr$(\hat \Theta, \hat \tau)=0$, so that we can obtain the performance depicted in the
top left-hand plot of Figure 3.

\medskip

\noindent \textbf{Example 3 (prior information about a 2-dimensional parameter vector, equi-tailed confidence interval
for $\boldsymbol{\theta}$)} \newline
Consider the model and parameter of interest $\theta$ described in the Introduction. Suppose that $p > 2$ and
$n-p$ is small.
Let the 2-dimensional parameter vector $\boldsymbol{\psi}$ be defined to be $\boldsymbol{C}^T \boldsymbol{\beta} - \boldsymbol{t}$,
where $\boldsymbol{C}$ is a specified $p \times 2$ matrix with linearly independent columns and $\boldsymbol{t}$ is a specified
2-vector. Suppose that $\boldsymbol{a}$ does not belong to the linear subspace spanned by the columns of $\boldsymbol{C}$.
Also suppose that previous experience with similar data sets and/or
expert opinion and scientific background suggest that $\boldsymbol{\psi} = \boldsymbol{0}$.
In other words, suppose that we have uncertain prior information that $\boldsymbol{\psi} = \boldsymbol{0}$.
Let $\hat{\Psi} = \boldsymbol{C}^T \hat{\boldsymbol{\beta}} - \boldsymbol{t}$.

Suppose that our aim is to find an {\sl equi-tailed} $1-\alpha$ confidence interval for
$\theta$ that utilizes this uncertain prior information. If Cov$(\hat \Theta, \hat{\Psi}_i)=0$ then we can find such
a confidence interval by letting $\tau = \psi_i$ ($i=1,2$). If, on the other hand,
Cov$(\hat \Theta, \hat{\Psi}_1) \ne 0$ and Cov$(\hat \Theta, \hat{\Psi}_2) \ne 0$ then we can find such a confidence
interval by letting
\begin{equation*}
\tau = \psi_1 - \frac{\text{Cov}(\hat \Theta, \hat{\Psi}_1)}{\text{Cov}(\hat \Theta, \hat{\Psi}_2)} \, \psi_2
\end{equation*}
and noting that Corr$(\hat \Theta, \hat \tau)=0$, where
\begin{equation*}
\hat{\tau} = \hat{\Psi}_1 - \frac{\text{Cov}(\hat \Theta, \hat{\Psi}_1)}{\text{Cov}(\hat \Theta, \hat{\Psi}_2)} \, \hat{\Psi}_2.
\end{equation*}

\medskip

\noindent {\sl Remark 5.4} \ \  As stated in Appendix A, we have chosen the functions $b$ and $s$ to be cubic
splines in the interval $[0,d]$. Other choices of parametric forms for these functions are also possible.
For example, one could choose these functions to be piecewise cubic Hermite interpolating polynomials in this
interval.

\medskip

\noindent {\sl Remark 5.5} \ \ Instead of minimizing the criterion \eqref{criterion} (subject to the coverage constraint)
one could minimize the following criterion (subject to the same coverage constraint)
\begin{equation}
\label{new_criterion}
\xi \int_{-\infty}^{\infty} (e(\gamma;s) - 1) \, d \gamma +
\int_{-\infty}^{\infty} (e(\gamma;s) - 1) \, \phi(\gamma;v) \, d \gamma
\end{equation}
where $\phi(\gamma;v)$ denotes the $N(0,v^2)$ probability density function and $v$ is a small
positive number. However, we expect that the use of \eqref{new_criterion} as an objective function
will lead to confidence intervals that are close to the corresponding confidence intervals obtained by using
\eqref{criterion} as the objective function.

\medskip

\noindent {\sl Remark 5.6} \ \ Instead of minimizing the criterion \eqref{criterion}, subject to the coverage constraint,
we may proceed as follows. We minimize $e(\gamma=0;s)$, subject to both this coverage constraint and the constraint
that $\max_{\gamma} e(\gamma;s) \le \ell$, where $\ell$ is specified number satisfying $\ell > 1$.
Theorems 1, 2 and 3 are relevant to this procedure. Also, the obvious analogue of Corollary 1 holds for this
procedure. The performance of the
confidence interval that results from this procedure improves as $|\text{Corr}(\hat \Theta, \hat \tau)|$
increases and $n-p$ decreases. Figure 5 shows the performance of the confidence interval resulting
from this procedure when $\text{Corr}(\hat{\Theta}, \hat{\tau}) =
0.816496$, $n-p=1$, $1-\alpha = 0.95$ and $\ell=1.0308$, so that $\max_{\gamma} e(\gamma;s)$ is the same as
in Figure 2.

\FloatBarrier

% The following command was used to prevent the figures moving around too much.
\FloatBarrier

\begin{figure}[h]
%\label{Figure2}
\vskip-0.5cm
    \hskip-0.5cm
    %\centering
    \includegraphics[scale=0.38]{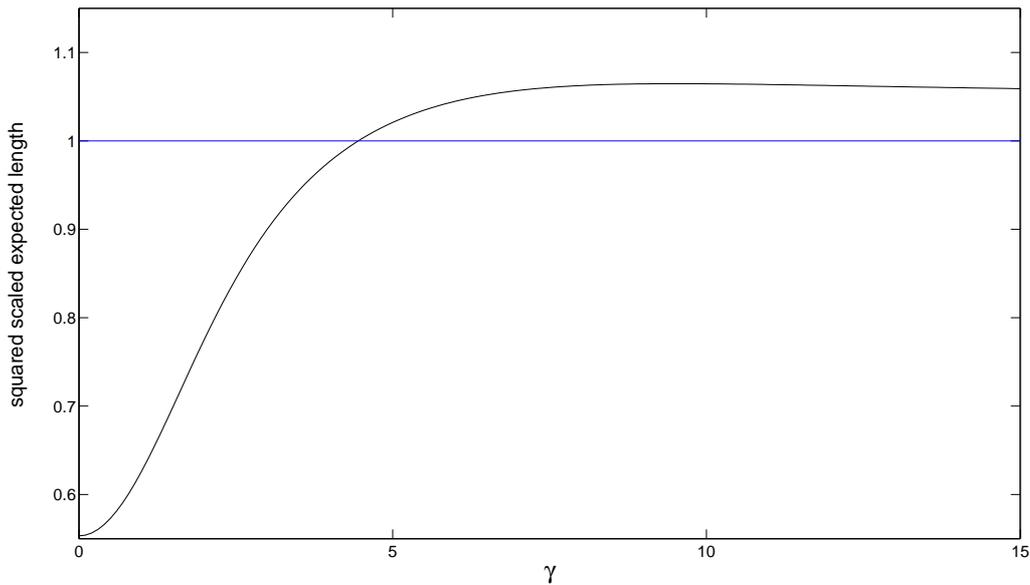}
    %\vskip-2.7cm
    \caption{Plot of the squared scaled expected length $e^2(\gamma;s)$ (as a function of
   $\gamma = \tau/(\sigma \sqrt{v_{22}})$) for the $1-\alpha$ confidence interval
for $\theta$ when $\text{Corr}(\hat{\Theta}, \hat{\tau}) =
0.816496$, $n-p=1$, $1-\alpha = 0.95$, $\ell=1.0308$, $d = 50$ and
the knots of the cubic splines $b$  and $s$ (in the interval $[0,d]$) are at $0, 15, 18, 21, 24, 27, 30, 50$ and
 at  $0, 2, 4, 6, 8, 10, 25, 50$, respectively.}
\end{figure}

\FloatBarrier

\medskip

\noindent {\sl Remark 5.7} \ \ In the example presented at the end of Section 2, the uncertain prior
information is that $\beta_{12}$, $\beta_{13}$ and $\beta_{23}$ are all zero. As noted in the description
of this example, this implies the uncertain prior information that $\tau = \beta_{23} - \beta_{13}$
is zero. By extending the work of
Kabaila \& Giri (2009a) to the case of uncertain prior information that a {\it vector} parameter is zero, it should
be possible (using the methods of Kabaila \& Farchione, 2012)
to construct a confidence interval for $\theta$ that utilizes the original prior information
(that $\beta_{12}$, $\beta_{13}$ and $\beta_{23}$ are all zero) more effectively.

\medskip

\begin{center}
{\bf 6. Conclusion}
\end{center}

%\medskip

Using computations and new theoretical results, we have shown that
the performance of the Kabaila \& Giri (2009a) confidence interval for $\theta$
improves as $|\text{Corr}(\hat \Theta, \hat \tau)|$ increases and $n-p$ decreases.
The improvement in performance of this confidence interval
as $|\text{Corr}(\hat \Theta, \hat \tau)|$ increases and $n-p$ decreases, is illustrated by
Figures 2, 3 and 4.

\bigskip

%\newpage
\begin{center}
{\bf {Appendix A: Computation of the KG confidence interval}}
\end{center}

\medskip

In addition to requiring that $b \in {\cal B}$ and $s \in {\cal S}$, we require that the
functions $b$ and $s$ are continuous.
For computational
tractability, $b$ and $s$ need to be restricted further. Kabaila \& Giri (2009a) take
$b$ and $s$ to be cubic splines in the interval $[0,d]$. We restrict the functions $b$ and
$s$ even further. We require the function $s$ to be {\sl unimodal} on the interval $[0,d]$. In other words,
we require that $s$ satisfies the condition that there exists $q \in (0,d)$ such that $s(x)$
is (a) a strictly increasing function of $x \in [0,q]$ and (b) a strictly decreasing function
of $x \in [q,d]$. If $\text{Corr}(\hat \Theta, \hat \tau) \ne 0$ then the function $b$ is also
required to be unimodal on the interval $[0,d]$.
Let ${\cal B}^*$ and ${\cal S}^*$ denote the subsets of ${\cal B}$ and ${\cal S}$,
respectively, that satisfy these requirements.

For judiciously chosen values of $d$, $\xi$ and the knots of the cubic splines for $b$ and $s$ in $[0,d]$,
we carry out the following computational procedure.

\medskip

\noindent {\sl Computational Procedure:} \ Compute $b \in {\cal B}^*$ and $s \in {\cal S}^*$ such that
(a) the minimum of the coverage probability $c(\gamma;b,s, \rho)$ over $\gamma$ is
$1-\alpha$ and (b) the criterion \eqref{criterion} is minimized.
Theorem 1 of Kabaila \& Giri (2009a) provides computationally
convenient expressions for $c(\gamma;b,s, \rho)$ and $e(\gamma;s)$.
Discussion 5.6 of this paper provides some further information about this computation.
A simplified expression for \eqref{criterion} is provided in Appendix B.
The resulting confidence interval is assessed using the following plots: plots of the functions
$b$ and $s$ on the interval $[0,d]$ and plots of the coverage probability $c(\gamma;b,s, \rho)$
the squared scaled expected length $e^2(\gamma;s)$, as functions of $\gamma \ge 0$.

\medskip

\noindent Based on these plots, we choose
$d$, $\xi$ and the knots of the cubic splines for $b$ and $s$ in $[0,d]$, so that the confidence
interval has not only desirable coverage probability and scaled expected length properties, but
also the functions $b$ and $s$ have desirable properties, such as smoothness. We refer to the resulting confidence interval
as the KG $1-\alpha$ confidence interval.

\bigskip

\begin{center}{\bf {Appendix B: Simplified expression for the criterion \eqref{criterion}}}
\end{center}

\medskip

In this appendix we provide a simplified expression for \eqref{criterion}.
Define $W = \hat \sigma/\sigma$.
Note that $W$ has the same distribution as $\sqrt{Q/(n-p)}$ where $Q
\sim \chi^2_{n-p}$.
Let $f_W$ denote the probability density function of $W$.
According to (8) of Kabaila \& Giri (2009a),
\eqref{criterion} is equal to
\begin{equation*}
\frac{2} {t(n-p) \, E(W)}
 \int^{\infty}_0 \int^{d}_{0} \left (s(x) - t(n-p) \right )
(\xi + \phi(w x)) \, dx \,  w^2 \,  f_W(w)  \, dw.
\end{equation*}
where $\phi$ denotes the $N(0,1)$ probability density function.
Now this is equal to
\begin{equation*}
\frac{2} {t(n-p) \, E(W)} \left ( \xi \int_0^d (s(x) - t(n-p)) \, dx +
\int_0^d \big (s(x) - t(n-p) \big) \, \int_0^{\infty} \phi(w x)  \,  w^2 \,  f_W(w)  \, dw
\, dx \right ).
\end{equation*}
By the following lemma, this is equal to
\begin{equation*}
\frac{2} {t(m) \, E(W)} \int_0^d \left (s(x) - t(m) \right) \,
\left (\xi + \frac{1}{\sqrt{2 \pi}} \left ( \frac{m}{x^2+m} \right )^{(m/2)+1} \right ) \,
dx,
\end{equation*}
where $m = n-p$.

\begin{lemma}
\begin{equation}
\label{lemma1_result}
\int_0^{\infty} \phi(w x)  \,  w^2 \,  f_W(w)  \, dw = \frac{1}{\sqrt{2 \pi}} \left ( \frac{m}{x^2+m} \right )^{(m/2)+1}.
\end{equation}
\end{lemma}

\begin{proof}

Note that $f_W(w) = 2 m w f_m(m w^2)$, where $f_m$ denotes the $\chi_m^2$ probability density function.
Substituting the expressions for $\phi$ and $f_W$ into the left hand side of \eqref{lemma1_result},
we find that this is equal to
\begin{equation*}
\frac{2 m^{(m/2)}}{\sqrt{2 \pi} \, \Gamma(m/2) \, 2^{m/2}}
\int_0^{\infty} w^{m+1} \exp \left(- \frac{1}{2}(m+x^2) w^2 \right ) \, dw
\end{equation*}
By (A2.1.3) of Box \& Tiao (1973), this is equal to the right hand side of \eqref{lemma1_result}.

\end{proof}

%\medskip

\newpage

\begin{center}
\noindent {\bf {Appendix C: Proof of Theorem 1}}
\end{center}
%\medskip

In this appendix, we prove Theorem 1. Suppose that Corr$(\hat \Theta, \hat \tau)=0$ and that the function $s \in {\cal S}$ is given.
Fix $\gamma \in \mathbb{R}$.

Maximizing $c(\gamma;b,s,\rho)$ with respect to $b \in {\cal B}$ is equivalent to minimizing
$1 - \alpha - c(\gamma;b,s,\rho)$ with respect to $b \in {\cal B}$. Define
\begin{align*}
k(x,w,\gamma) &= \Phi \big(b(x) w + s(|x|) w \big) - \Phi \big(b(x) w - s(|x|) w \big) \\
k^{\dag}(w) &= 2 \Phi \big(t(n-p) w \big) - 1,
\end{align*}
where $\Phi$ denotes the $N(0,1)$ distribution function. According to p.307 of Kabaila, Giri and Leeb (2010),
\begin{equation*}
1 - \alpha - c(\gamma;b,s,\rho) = - \big (r_1(b,s,\gamma) + r_2(b,s,\gamma) \big)
\end{equation*}
where
\begin{align*}
r_1(b,s,\gamma) &= \int_0^{\infty} \int_0^d \big (k(x,w,\gamma)-k^{\dag}(w) \big) \, \phi(wx-\gamma) \, w \, f_W(w) \, dx \, dw \\
r_2(b,s,\gamma) &= \int_0^{\infty} \int_0^d \big (k(-x,w,\gamma)-k^{\dag}(w) \big) \, \phi(wx+\gamma) \, w \, f_W(w) \, dx \, dw.
\end{align*}
Thus, minimizing $1 - \alpha - c(\gamma;b,s,\rho)$ with respect to $b \in {\cal B}$ is equivalent to maximizing
$r_1(b,s,\gamma) + r_2(b,s,\gamma)$ with respect to $b \in {\cal B}$.

According to p.309 of Kabaila, Giri \& Leeb (2010), for fixed $s>0$ and $w>0$,
$\Phi(bw+sw)-\Phi(bw-sw)$ is maximized with respect to $b \in \mathbb{R}$ at $b=0$. Thus
$\Phi \big(b(x) w + s(x) w \big) - \Phi \big(b(x) w - s(x) w \big)$ is, for each $x \in [0,d]$ and $w>0$, maximized with respect
to $b(x) \in \mathbb{R}$ at $b(x)=0$. Since $\phi(w x - \gamma) w f_W(w) > 0$ for all $x \in [0,d]$ and $w>0$,
$r_1(b,s,\gamma)$ is maximized with respect to the function $b \in {\cal B}$ by setting $b \equiv 0$.
A similar argument shows that $r_2(b,s,\gamma)$ is maximized with respect to the function $b \in {\cal B}$ by setting $b \equiv 0$.
Thus, $r_1(b,s,\gamma) + r_2(b,s,\gamma)$ is maximized with respect to the function $b \in {\cal B}$ by setting $b \equiv 0$.

\medskip

\begin{center}
\noindent {\bf {Appendix D: Proof of Corollary 1}}
\end{center}

Suppose that Corr$(\hat \Theta, \hat \tau)=0$. Suppose that ${\cal B}^*$ is a subset of ${\cal B}$ that includes the function
$b \equiv 0$.
Also suppose that ${\cal S}^*$ is a subset of ${\cal S}$.

The infimum over $(b,s) \in {\cal B}^* \times {\cal S}^*$ of  \eqref{criterion},
subject to the constraint \eqref{cov_constr}, is less than or equal
to the infimum over $s \in {\cal S}^*$ of \eqref{criterion}, subject to this constraint, when $b \equiv 0$.
We complete the proof by contradiction. Suppose that the infimum over $(b,s) \in {\cal B}^* \times {\cal S}^*$ of  \eqref{criterion},
subject to the constraint \eqref{cov_constr}, is less than
to the infimum over $s \in {\cal S}^*$ of \eqref{criterion}, subject to this constraint, when $b \equiv 0$.
Thus there exists $(b^{\prime},s^{\prime}) \in {\cal B}^* \times {\cal S}^*$
such that the constraint \eqref{cov_constr}, evaluated at $(b,s)=(b^{\prime},s^{\prime})$, is satisfied and
\eqref{criterion}, evaluated at $(b,s)=(b^{\prime},s^{\prime})$, is less than
the infimum over $s \in {\cal S}^*$ of \eqref{criterion}, subject to this constraint, when $b \equiv 0$.

By Theorem 1, the following is true. If we let $b \equiv 0$ then $(b,s)=(b,s^{\prime})$ satisfies the constraint \eqref{cov_constr}.
Also,
\eqref{criterion}, evaluated at $(b,s)=(b,s^{\prime})$, is equal to \eqref{criterion}, evaluated at
$(b,s)=(b^{\prime},s^{\prime})$. We have established a contradiction.

\medskip

\begin{center}
\noindent {\bf {Appendix E: Proof of Theorem 2}}
\end{center}

In this appendix, we prove Theorem 2.
Suppose that Corr$(\hat \Theta, \hat \tau)=0$ and that $b \equiv 0$.
The confidence interval $J(b,s)$ has the form \eqref{ci_b_0}.
Let $G = (\hat{\Theta} - \theta)/(\sigma \sqrt{v_{11}})$ and $H = \hat{\tau}/(\sigma \sqrt{v_{22}})$.
Note that $G$ and $H$ are independent random variables and $G \sim N(0,1)$.
Now
\begin{equation}
\label{less_low}
P \left ( \theta < \hat \Theta \,- \,
\sqrt{v_{11}} \hat \sigma \, s \left(\frac{|\hat{\tau}|}{\hat \sigma \sqrt{v_{22}}} \right ) \right )
= P \left ( G > W s \left( \frac{|H|}{W} \right) \right )
\end{equation}
Also,
\begin{equation}
\label{greater_up}
P \left ( \theta > \hat \Theta \,+ \,
\sqrt{v_{11}} \hat \sigma \, s \left(\frac{|\hat{\tau}|}{\hat \sigma \sqrt{v_{22}}} \right ) \right )
= P \left ( \tilde{G} > W s \left( \frac{|H|}{W} \right) \right )
\end{equation}
where $\tilde{G} = -G$. Thus \eqref{less_low} = \eqref{greater_up}.

\medskip

\begin{center}
{\bf {Appendix F: Proof of Theorem 3}}
\end{center}

Suppose that Corr$(\hat \Theta, \hat \tau)=0$ and that $b \equiv 0$.
Theorem 3 provides a lower bound for $e(\gamma=0;s)-1$, subject to the constraints
that $s \in {\cal S}$ and $c(\gamma;b,s,\rho) \ge 1-\alpha$ for all $\gamma$.
We prove this result using the framework of compromise decision theory (Kempthorne, 1983, 1987, 1988).
Specifically, we use Theorem 2.2 (a) of Kabaila \& Tuck (2008) to prove this result.

Define $R_1(s;\gamma) = e(\gamma;s) - 1$. Also define $\pi_1$ to be the unit step
function. Thus
\begin{equation*}
\int_{-\infty}^{\infty} R_1(s;\gamma) \, d \pi_1(\gamma) = e(\gamma=0;s) - 1.
\end{equation*}
Now define $R_2(s;\gamma) = 1 - \alpha - c(\gamma;b,s,\rho)$. Define $\pi_2$ to the
unit step function. Now define
\begin{equation*}
g(s;\lambda) = \lambda \int_{-\infty}^{\infty} R_1(s;\gamma) \, d \pi_1(\gamma)
+ (1-\lambda) \int_{-\infty}^{\infty} R_2(s;\gamma) \, d \pi_2(\gamma),
\end{equation*}
where $0 < \lambda < 1$. Let $m=n-p$.
For each positive integer $m$, we will define $\lambda(m) \in (0,1)$
and we will find $s$ that minimizes
$g(s;\lambda(m))$ with respect to $s \in {\cal S}$. Denote this minimizing value
of $s$ by $s_{\lambda(m)}$.  We will also note that
\begin{equation*}
\sup_{\gamma} R_2(s_{\lambda(m)};\gamma) =0
\end{equation*}
and that
\begin{equation}
\label{def_nu_m}
\nu_m = \sup_{\gamma} R_2(s_{\lambda(m)};\gamma) - \int_{-\infty}^{\infty} R_2(s_{\lambda(m)};\gamma) \, d \pi_2(\gamma)
\end{equation}
converges to 0 as $m \rightarrow \infty$. Theorem 2.2 (a) of Kabaila \& Tuck (2008) implies that
\begin{equation*}
\inf_{s \in \tilde{\cal S}} \int_{-\infty}^{\infty} R_1(s;\gamma) \, d \pi_1(\gamma) \ge
\int_{-\infty}^{\infty} R_1(s_{\lambda(m)};\gamma) \, d \pi_1(\gamma) - \frac{1-\lambda(m)}{\lambda(m)} \nu_m,
\end{equation*}
for each positive integer $m$. In other words,
\begin{equation}
\label{almost_there}
\inf_{s \in \tilde{\cal S}} \, e(\gamma=0;s) - 1 \ge
e(\gamma=0;s_{\lambda(m)}) - 1  - \frac{1-\lambda(m)}{\lambda(m)} \nu_m.
\end{equation}
We will then note that $e \big(\gamma=0;s_{\lambda(m)} \big) \ge 1$ and
show that $\nu_m (1-\lambda(m))/\lambda(m)$ converges to 0, as $m \rightarrow \infty$.

It follows from Theorem 1 (b) of Kabaila \& Giri (2009a) that
\begin{equation*}
e(\gamma=0;s) - 1 = \frac{2}{t(m)E(W)} \int_0^d \big ( s(x) - t(m) \big)
\int_0^{\infty} \phi(w x) \, w^2 \, f_W(w) \, dw \, dx,
\end{equation*}
where $\phi$ denotes the $N(0,1)$ probability density function.
It follows from p.307 of Kabaila, Giri \& Leeb (2010) that $1-\alpha-c(\gamma;b,s,\rho)$ is equal to
\begin{equation*}
-2 \int_0^d \int_0^{\infty} \big( \Phi(s(x) w) - \Phi(t(m) w) \big)
\big( \phi(w x - \gamma) + \phi(w x + \gamma) \big) w \, f_W(w) \, dw \, dx,
\end{equation*}
where $\Phi$ denotes the $N(0,1)$ distribution function.
Thus
\begin{align*}
g(s;\lambda) = &\lambda \frac{2}{t(m)E(W)} \int_0^d \big ( s(x) - t(m) \big) \int_0^{\infty} \phi(w x) \, w^2 \, f_W(w) \, dw \, dx \\
&-4 (1-\lambda) \int_0^d \int_0^{\infty} \big( \Phi(s(x) w) - \Phi(t(m) w) \big) \,
\phi(w x ) \, w \, f_W(w) \, dw \, dx.
\end{align*}
Minimizing this function with respect to $s \in {\cal S}$ is equivalent to minimizing
\begin{align*}
\tilde{g}(s;\lambda) =  \int_0^d \bigg (&\frac{\lambda}{t(m)E(W)}  \int_0^{\infty} \phi(w x) \, w^2 \, f_W(w) \, dw \,  s(x)  \\
&-2 (1-\lambda) \int_0^{\infty} \Phi(s(x) w) \,
\phi(w x ) \, w \, f_W(w) \, dw \bigg ) dx
\end{align*}
with respect to $s \in {\cal S}$. We find a minimizing value of $s \in {\cal S}$ as follows. For each $x \in [0,d)$, we minimize
\begin{equation}
\label{criterion_to_minimize}
\left (\frac{\lambda}{2 (1-\lambda) t(m) E(W)}  \int_0^{\infty} \phi(w x) \, w^2 \, f_W(w) \, dw \right)  t
-\int_0^{\infty} \Phi(t w) \,
\phi(w x ) \, w \, f_W(w) \, dw
\end{equation}
with respect to $t>0$ and then set $s(x)$ equal to this minimizing value.
The derivative of \eqref{criterion_to_minimize} with respect to $t$ is equal to
\begin{equation}
\label{deriv_criterion_to_minimize}
\frac{\lambda}{2 (1-\lambda) t(m) E(W)}  \int_0^{\infty} \phi(w x) \, w^2 \, f_W(w) \, dw
-\int_0^{\infty} \phi(t w) \,
\phi(w x ) \, w^2 \, f_W(w) \, dw.
\end{equation}
We simplify this expression using the following lemma.

\begin{lemma}
\begin{equation*}
\int_0^{\infty} \phi(tw) \phi(w x) w^2 f_W(w) \, dw = \frac{1}{2 \pi}
\left ( \frac{m}{t^2+x^2+m} \right)^{(m/2)+1}
\end{equation*}

\end{lemma}

\begin{proof}
Note that
\begin{equation*}
\phi(tw) \phi(w x) = \frac{1}{\sqrt{2 \pi}} \phi(w \tilde{x}),
\end{equation*}
where $\tilde{x} = \sqrt{t^2 + x^2}$. Hence
\begin{align*}
\int_0^{\infty} \phi(tw) \phi(w x) w^2 f_W(w) \, dw
&= \frac{1}{\sqrt{2 \pi}} \int_0^{\infty} \phi(w \tilde{x}) w^2 f_W(w) \, dw \\
&=\frac{1}{2 \pi}
\left ( \frac{m}{t^2+x^2+m} \right)^{(m/2)+1}
\end{align*}
by Lemma 1.

\end{proof}

By this lemma and Lemma 1 (stated in Appendix B), \eqref{deriv_criterion_to_minimize} is equal to
\begin{equation}
\label{simpl_deriv_criterion_to_minimize}
\frac{1}{2 \pi} \left ( \frac{\lambda}{(1-\lambda) t(m) E(W)} \sqrt{\frac{\pi}{2}}
\left ( \frac{m}{x^2+m} \right )^{(m/2)+1}
- \left ( \frac{m}{t^2+x^2+m} \right )^{(m/2)+1}  \right ).
\end{equation}
This is an increasing function of $t$, that approaches a positive number as $t \rightarrow \infty$.
Define $\lambda(m)$ to be the solution for $\lambda \in (0,1)$ of
\begin{equation*}
\sqrt{m} \sqrt{ \left ( \sqrt{ \frac{2}{\pi}} \frac{(1-\lambda) t(m) E(W)}{\lambda} \right )^{1/((m/2)+1)} - 1}
= t(m).
\end{equation*}
Henceforth, suppose that $\lambda = \lambda(m)$. Note that \eqref{simpl_deriv_criterion_to_minimize}
approaches a negative number as $t \downarrow 0$. Thus, for each $x \in [0,d)$,
we find the value of $t>0$ that minimizes
\eqref{criterion_to_minimize} by solving \eqref{simpl_deriv_criterion_to_minimize}=0 for $t>0$.
For each $x \in [0,d)$, this solution is
$t = \sqrt{1+(x^2/m)} \, t(m)$. Thus
\begin{equation*}
s_{\lambda(m)}(x) =
\begin{cases}
\sqrt{\displaystyle 1+\frac{x^2}{m}} \, t(m) &\text{ for } x \in [0,d) \\
t(m) &\text{ for } x \ge d.
\end{cases}
\end{equation*}

Now
\begin{equation*}
\sup_{\gamma} R_2(s_{\lambda(m)}; \gamma) = 1 - \alpha - \inf_{\gamma} c(\gamma; b \equiv 0, s_{\lambda(m)}, \rho=0).
\end{equation*}
Since $s_{\lambda(m)}(x) \ge t(m)$ for all $x \ge 0$, the following easily-proved lemma implies that
\begin{equation}
\label{sup_ineq}
\sup_{\gamma} R_2 (s_{\lambda(m)}; \gamma ) \le 0.
\end{equation}

\begin{lemma}

Suppose that $b: \mathbb{R} \rightarrow \mathbb{R}$,
$s: [0, \infty) \rightarrow (0, \infty)$ and $\tilde{s}: [0, \infty) \rightarrow (0, \infty)$ are measurable functions.
Also suppose that
$\tilde{s}(x) \ge s(x)$ for all $x \ge 0$. Then $c(\gamma;b,\tilde{s},\rho) \ge c(\gamma;b,s,\rho)$
for all $\gamma$.

\end{lemma}

\noindent The following lemma implies that
$c(\gamma; b \equiv 0, s_{\lambda(m)}, \rho=0) \rightarrow 1-\alpha$, as $\gamma \rightarrow \infty$.
It follows from \eqref{sup_ineq} that
\begin{equation*}
\sup_{\gamma} R_2(s_{\lambda(m)}; \gamma) = 0.
\end{equation*}

\begin{lemma}

Suppose that the positive integer $m$, $b \in {\cal B}$, $s \in {\cal S}$ and $\rho \in (-1,1)$ are given.
Then $c(\gamma; b, s, \rho) \rightarrow 1-\alpha$, as $\gamma \rightarrow \infty$.

\end{lemma}

\begin{proof}
It is an immediate consequence of a result stated on p.3428 of Kabaila \& Giri (2009a) that
\begin{equation*}
\big |c(\gamma; b, s, \rho) - (1-\alpha) \big | \le \int_0^{\infty} \int_{-dw}^{dw} \phi(h-\gamma) \, dh \, f_W(w) \, dw
\end{equation*}
where $f_W$ denotes the probability density function of $W = \hat \sigma/\sigma$. The result is a straightforward
consequence of this inequality.

\end{proof}

Define $\nu_m$ by \eqref{def_nu_m} and note that
\begin{equation*}
\nu_m = c \big(\gamma=0; b \equiv 0, s_{\lambda(m)}, \rho=0 \big) - (1-\alpha).
\end{equation*}
By Lemma 3,
\begin{equation*}
c \big(\gamma=0; b \equiv 0, s_{\lambda(m)}, \rho=0 \big) \le c \Big(\gamma=0; b \equiv 0, s \equiv \sqrt{1+(d^2/m)} t(m), \rho=0 \Big)
\end{equation*}
where $s \equiv \sqrt{1+(d^2/m)} t(m)$ denotes the function $s$ that satisfies
$s(x)=\sqrt{1+(d^2/m)} t(m)$ for all $x \in \mathbb{R}$. Thus $\nu_m \downarrow 0$ as $m \rightarrow \infty$.
As noted earlier, \eqref{almost_there} holds. Since $e \big(\gamma=0;s_{\lambda(m)} \big) \ge 1$,
\begin{equation*}
\inf_{s \in \tilde{\cal S}} \, e(\gamma=0;s) \ge
1  - \frac{1-\lambda(m)}{\lambda(m)} \nu_m.
\end{equation*}
It may be shown that $\lim_{m \rightarrow \infty} \lambda(m)$ exists and belongs to $(0,1)$.
Thus, $\nu_m (1-\lambda(m))/\lambda(m) \rightarrow 0$, as $m \rightarrow \infty$.

\medskip

\begin{center}
\noindent {\large{\sl References}}
\end{center}

\rf BERGER, J. (1980). A robust generalized Bayes estimator and confidence region for a multivariate normal mean. {\sl Annals of Statistics} {\bf 8}, 716--761.

\smallskip

\rf BICKEL, P.J. (1984). Parametric robustness: small biases can be
worthwhile. {\sl Annals of Statistics} {\bf 12}, 864--879.

\smallskip

\rf BOX, G.E.P. \& TIAO, G.C. (1973). {\sl Bayesian Inference in Statistical Analysis}. New York: Wiley.

\smallskip

\rf BROWN, L.D., CASELLA, G. \& HWANG, J.T.G. (1995). Optimal confidence sets, bioequivalence and the Limacon of Pascal.
{\sl Journal of the American Statistical Association} {\bf 90}, 880--889.

\smallskip

\rf CASELLA, G. \& HWANG, J.T. (1983). Empirical Bayes confidence sets for the mean of a multivariate normal distribution. {\sl Journal of the American Statistical Association} {\bf 78}, 688--698.

\smallskip

\rf EFRON, B. (2006) Minimum volume confidence regions for a multivariate normal mean. {\sl Journal of the Royal Statistical Society,  Series B} {\bf 68}, 655--670.

\smallskip

\rf FARCHIONE, D. \& KABAILA, P. (2008). Confidence intervals for the normal mean utilizing prior information.
{\sl Statistics \& Probability Letters} {\bf 78}, 1094--1100.

\smallskip

\rf GOUTIS, C. \& CASELLA, G. (1991). Improved invariant confidence intervals for a normal variance.
{\sl Annals of Statistics} {\bf 19}, 2015--2031.

\smallskip

\rf HINKELMANN, K. \& KEMPTHORNE, O. (1994). Design and Analysis of Experiments, revised edition.
New York: John Wiley.

\smallskip

\rf HODGES, J.L. \& LEHMANN, E.L. (1952). The use of previous
experience in reaching statistical decisions. {\sl Annals of
Mathematical Statistics} {\bf 23}, 396--407.

\smallskip

\rf KABAILA, P. (1998). Valid confidence intervals in regression after variable selection.
{\sl Econometric Theory} {\bf 14}, 463--482.

\smallskip

\rf KABAILA P. (2009). The coverage properties of confidence regions after model
selection. {\sl International Statistical Review} {\bf 77},405--414.

\smallskip

\rf KABAILA, P. \& TUCK, J. (2008). Confidence intervals utilizing prior information in the Behrens-Fisher
problem. {\sl Australian \& New Zealand Journal of Statistics} {\bf 50}, 309--328.

\smallskip

\rf KABAILA, P. \& GIRI, K. (2009a). Confidence intervals in regression utilizing prior information.
{\sl Journal of Statistical Planning and Inference} {\sl 139}, 3419--3429.

\smallskip

\rf KABAILA, P. \& GIRI, K. (2009b). Upper bounds on the minimum coverage probability of confidence
intervals in regression after model selection. {\sl Australian \& New Zealand Journal of Statistics}
{\bf 51}, 271--288.

%\rf KABAILA, P. \& GIRI, K. (2009c). Large-sample confidence intervals %for the treatment difference in a two-period
%crossover trial, utilizing prior information. {\sl Statistics \& %Probability Letters} {\bf 79}, 652--658.

\smallskip

\rf KABAILA, P., GIRI, K. \& LEEB, H. (2010). Admissibility of the usual confidence interval in linear
regression. {\sl Electronic Journal of Statistics} {\bf 4}, 300--312.

\smallskip

\rf KABAILA, P. \& FARCHIONE, D. (2012). The minimum coverage probability of confidence intervals in
regression after a preliminary F test. {\sl Journal of Statistical Planning and Inference} {\bf 142}, 956--964.

\smallskip

\rf KEMPTHORNE, P.J. (1983). Minimax-Bayes compromise estimators. In
{\sl 1983 Business and Economic Statistics Proceedings of the
American Statistical Association}, Washington DC, pp.568--573.

\smallskip

\rf KEMPTHORNE, P.J. (1987).  Numerical specification of
discrete least favourable prior distributions.  {\sl SIAM
Journal on Scientific and Statistical Computing} {\bf 8}, 171--184.

\smallskip

\rf KEMPTHORNE, P.J. (1988). Controlling risks under different loss
functions: the compromise decision problem. {\sl Annals of
Statistics} {\bf 16}, 1594--1608.

\smallskip

\rf MAATTA, J.M. \& CASELLA, G. (1990). Decision-theoretic estimation. {\sl Statistical
Science} {\bf 5}, 90--120.

\smallskip

\rf MEAD, R. (1988). {\sl The Design of Experiments.}
Cambridge: Cambridge University Press.

\smallskip

\rf PRATT, J.W. (1961). Length of confidence intervals. {\sl Journal of
the American Statistical Association} {\bf 56}, 549--657.

\smallskip

\rf PUZA, B. \& O'NEILL, T. (2006a). Generalised Clopper-Pearson confidence intervals for the binomial proportion.
{\sl Journal of Statistical Computation and Simulation} {\bf 76}, 489 -- 508.

\smallskip

\rf PUZA, B. \& O'NEILL, T. (2006b). Interval estimation via tail functions.
{\sl Canadian Journal of Statistics} {\bf 34}, 299 -- 310.

\smallskip

\rf SALEH, A.K.Md.E. (2006). {\sl Theory of Preliminary Test and Stein-Type Estimation and Applications.}
Hoboken, NJ: Wiley.

\smallskip

\rf STEIN, C.M. (1962). Confidence sets for the mean of a multivariate normal distribution. {\sl Journal of the Royal Statistical Society,  Series B} {\bf 24}, 265--296.

\smallskip

\rf TSENG, Y-L. \& BROWN, L.D. (1997). Good exact confidence sets for a multivariate normal mean. {\sl Annals of Statistics} {\bf 25}, 2228--2258.

\end{document}